\documentclass[5p,twocolumn,times,number]{elsarticle}

\usepackage{graphicx}
\usepackage{amsmath}   

\begin{document}

\begin{frontmatter}

\title{The KLOE-2 High Energy Tagger Detector}

%\corref{cor}
%\ead{moricciani@roma2.infn.it}
%\cortext[cor]{Corresponding author}
\author[add1]{D.~Babusci}
\author[add1]{F.~Gonnella}
\author[add1]{L.~Iafolla}
\author[add1]{M.~Iannarelli}
\author[add2,add3]{M.~Mascolo}
\author[add2,add3]{R.~Messi}
\author[add3]{D.~Moricciani}
\author[add1]{A.~Saputi}
\author[add1]{E.~Turri}

\address[add1]{INFN-LNF, Frascati, Italy}
\address[add2]{Physics Dep."Tor Vergata" University, Roma, Italy}
\address[add3]{INFN-Roma "Tor Vergata", Rome, Italy}

\begin{abstract}
In order to fully reconstruct to the reaction $e^+e^- \to e^+e^- \gamma\gamma$ in the energy
region of the $\phi$ meson production, new detectors along the (DA$\Phi$NE) 
beam line have to be installed in order to detect the scattered $e^+e^-$.
The High Energy Tagger (HET) detector measures the deviation of leptons from 
their main orbit by determining their position and timing so to tag 
$\gamma\gamma$ physics events and disentangle them 
from background. 
The HET detectors are placed at the exit of the DA$\Phi$NE dipole magnets, 
11 m away from the IP, both on positron and electron lines. The HET
sensitive area is made up of a set of 28 plastic scintillators.
A dedicated DAQ electronics board based on a Xilinx Virtex-5 FPGA have
been developed for this detector. It provides a MultiHit TDC with a 
time resolution of the order of 500 ps and the possibility to  
acquire data any 2.5 ns, thus allowing to clearly identify the
correct bunch crossing.
First results of the commissioning run are presented.
\end{abstract}


\end{frontmatter}

\section{HET Detector}
\begin{figure}[h] 
	\begin{center}                               
		\includegraphics[width=.5\textwidth]{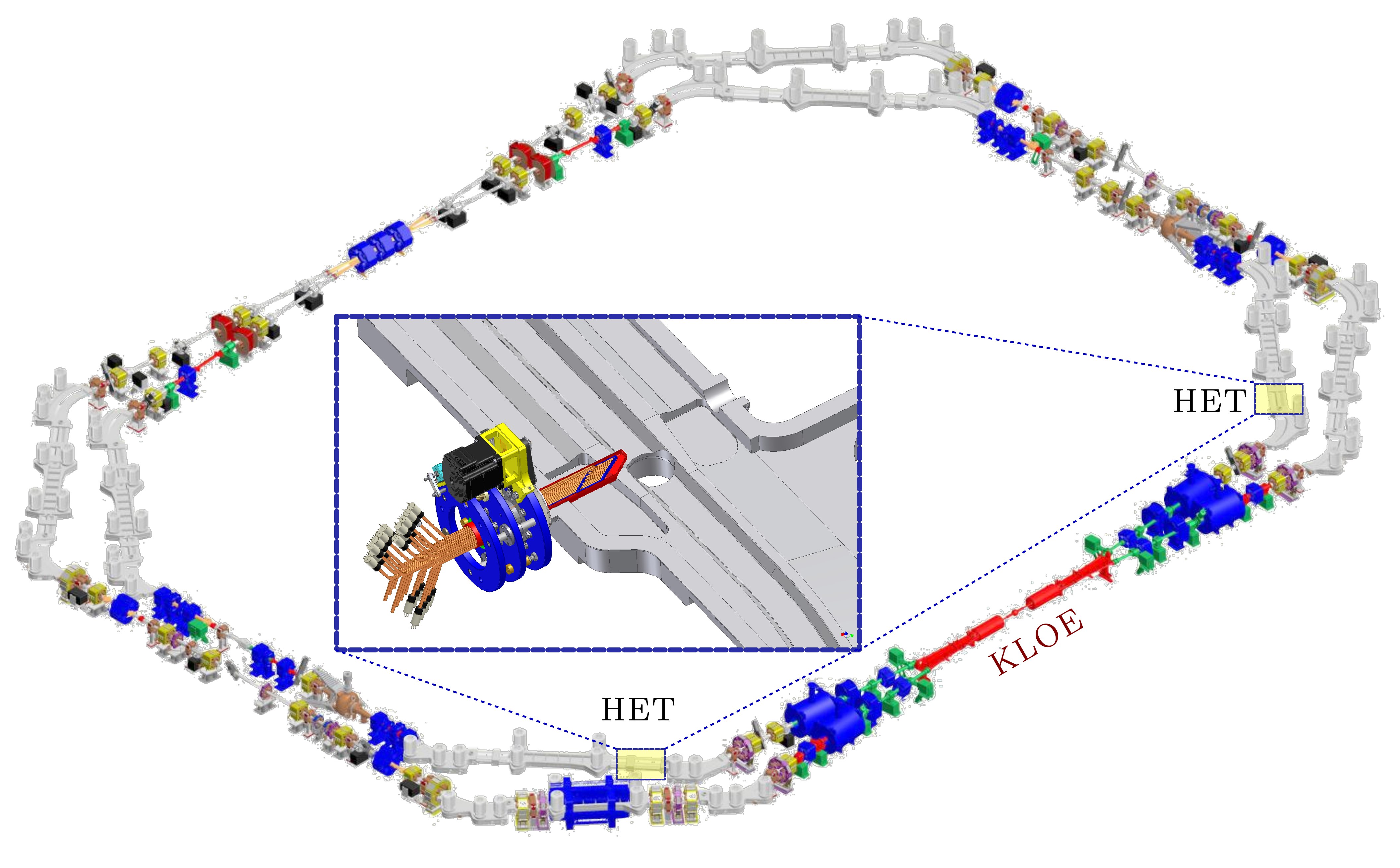}                
		\caption[The location of HET detectors on DA$\Phi$NE]{A drawing of the two HET detectors placed on DA$\Phi$NE lattice.}
			\label{fig:taggers}                          
	\end{center}                                 
\end{figure}

The High Energy Tagger (HET) detector now installed in KLOE-2 \cite{kloe2} is a 
position detector used for measuring the deviation of leptons from 
their main orbit in DA$\Phi$NE. By means of 
this measurement and of its timing, we are able to disentangle, and 
therefore to \emph{tag} $\gamma\gamma$ physics \cite{pi0} events.
Two HET detectors are placed at the exit of the 
dipole magnets (see Fig. \ref{fig:taggers}), {11} {m} away from the 
IP, both on positron and electron arm.
The sensitive area of the HET detector is made up of a set of 28 
plastic scintillators. The dimensions of each of them are $(\rm{3} 
\times \rm{5} \times \rm{6})~mm^3$.
One additional scintillator, of dimensions: $(\rm{3} \times \rm{50} 
\times \rm{6})~mm^3$ is used for coincidence purposes.
The light emitted by each of the 28 scintillators is read out 
through a plastic light guide by a photomultiplier. The 28 scintillators 
are placed at different distances from the beam-line, in such a way that 
the measurement of the distance, between the hitting particle and the beam, 
can be performed simply knowing which scintillator has been fired.

\begin{figure}[h] 
	\begin{center}                               
		\includegraphics[trim = 0mm 12cm 0mm 16cm, clip, width=.4\textwidth]{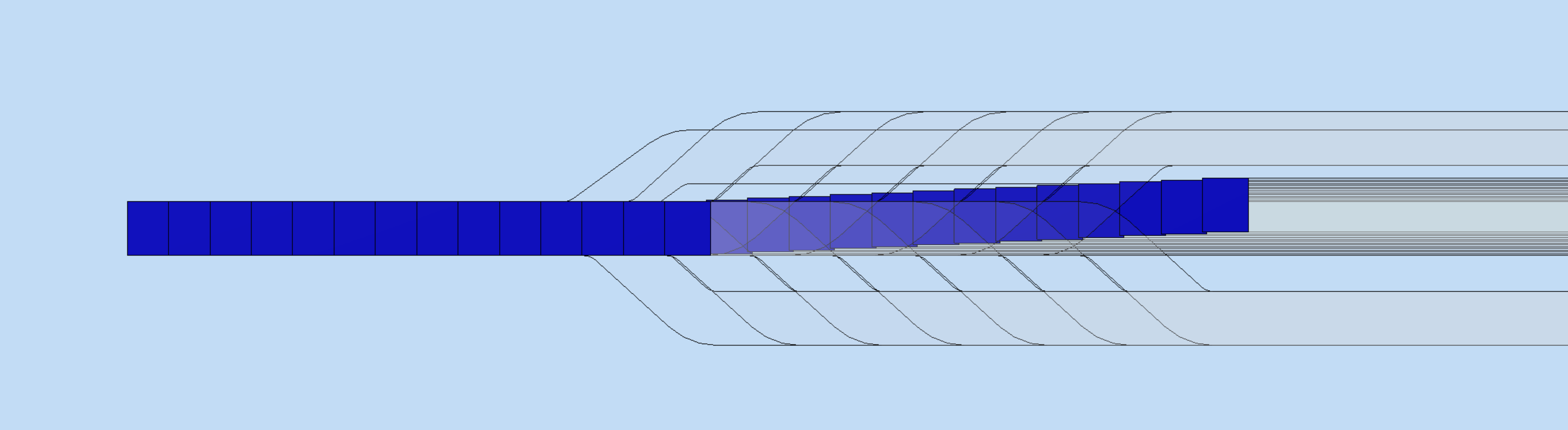}                
		\caption[Vertical profile of the HET scintillators]{A drawing of the vertical profile of the HET 
		scintillators (blue) with their light guides (transparent). The beam line is on the left of the drawing. 
		One can notice the vertical curvature used to follow the impinging position of the leptons, positrons in this case.}
		\label{fig:profilo}                          
	\end{center}                                 
\end{figure}

On the vertical plane (see Fig. \ref{fig:profilo}), the scintillators are aligned, 
with the exception of a slight curvature at big distance from the beam, this 
is due to maximize the HET efficiency due to the KLOE magnetic axial field 
and DA$\Phi$NE compensator. 
They show their $(\rm{5} \times \rm{6})~mm^2$ face to the impinging particles 
that go through them along the thickness of \rm{3 mm}. The scintillators are 
not placed side by side, on the contrary there is an overlap of \rm{0.5 mm} on 
the \rm{5 mm} side.

The plastic scintillator used is the EJ-228 premium plastic scintillator 
produced by Eljen Technology. It is intended for use in ultra-fast timing and 
ultra-fast counting applications and it is recommended for use in small sizes 
(any dimension less than \rm{100 mm}). The EJ-228 scintillator is composed 
of Polyvinyltoluene and has a density of \rm{1.023 g/cm$^3$} and a refractive 
index of 1.58 and a light output 67 $\%$ of anthracene, the emission spectra 
is peaked around 391 nm.

Because of the small dimensions of the scintillator in use, the total 
light yield, due to a crossing electron or positron, is quite small. For this reason, we 
chose a high quantum efficiency photomultiplier to minimize the 
probability of a particle, which crosses the scintillator, to go undetected.
The photomultipliers used are compact size and high quantum efficiency ones,
model R9880U-110 SEL produced by Hamamatsu Photonics. 
The quantum efficiency is about 35\% for a wavelength going in the range 
from \rm{300~nm} to \rm{400~nm}, well matching the EJ-228 emission. 

\begin{figure}[h] 
\centering
		\includegraphics[width=0.35\textwidth, trim= 0 20mm 0 20mm, clip]{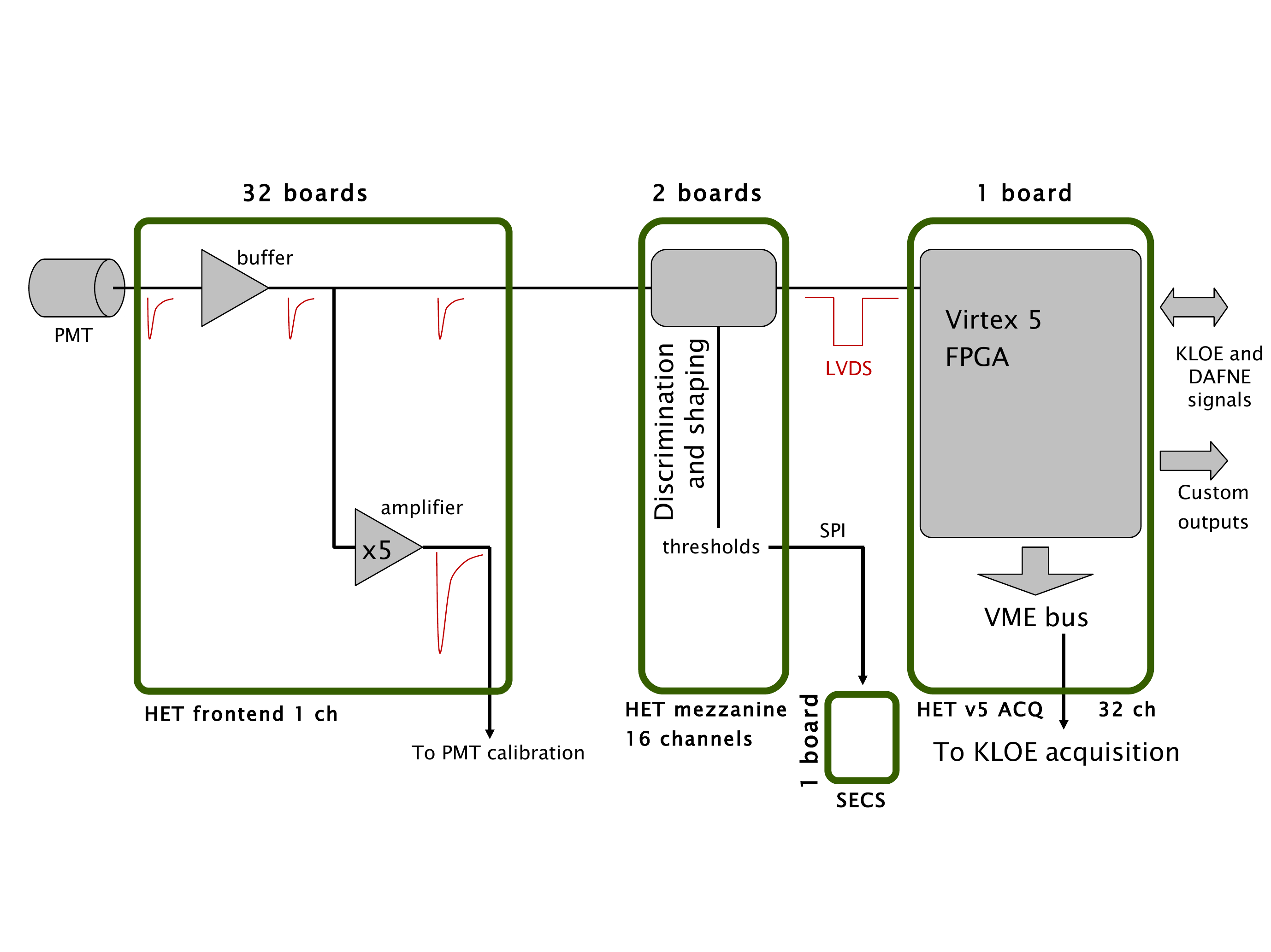}                
		\caption[Scheme of the HET electronic acquisition chain]{A scheme of the electronic 
		aquisition chain of the HET detector.}
		\label{fig:ELscheme}                          	
\end{figure}

\section{HET Electronics}

\begin{figure}[htpb] 
\centering	
		\includegraphics[angle=270, trim= 8mm 25mm 8mm 25mm, clip, width= 0.35\textwidth]{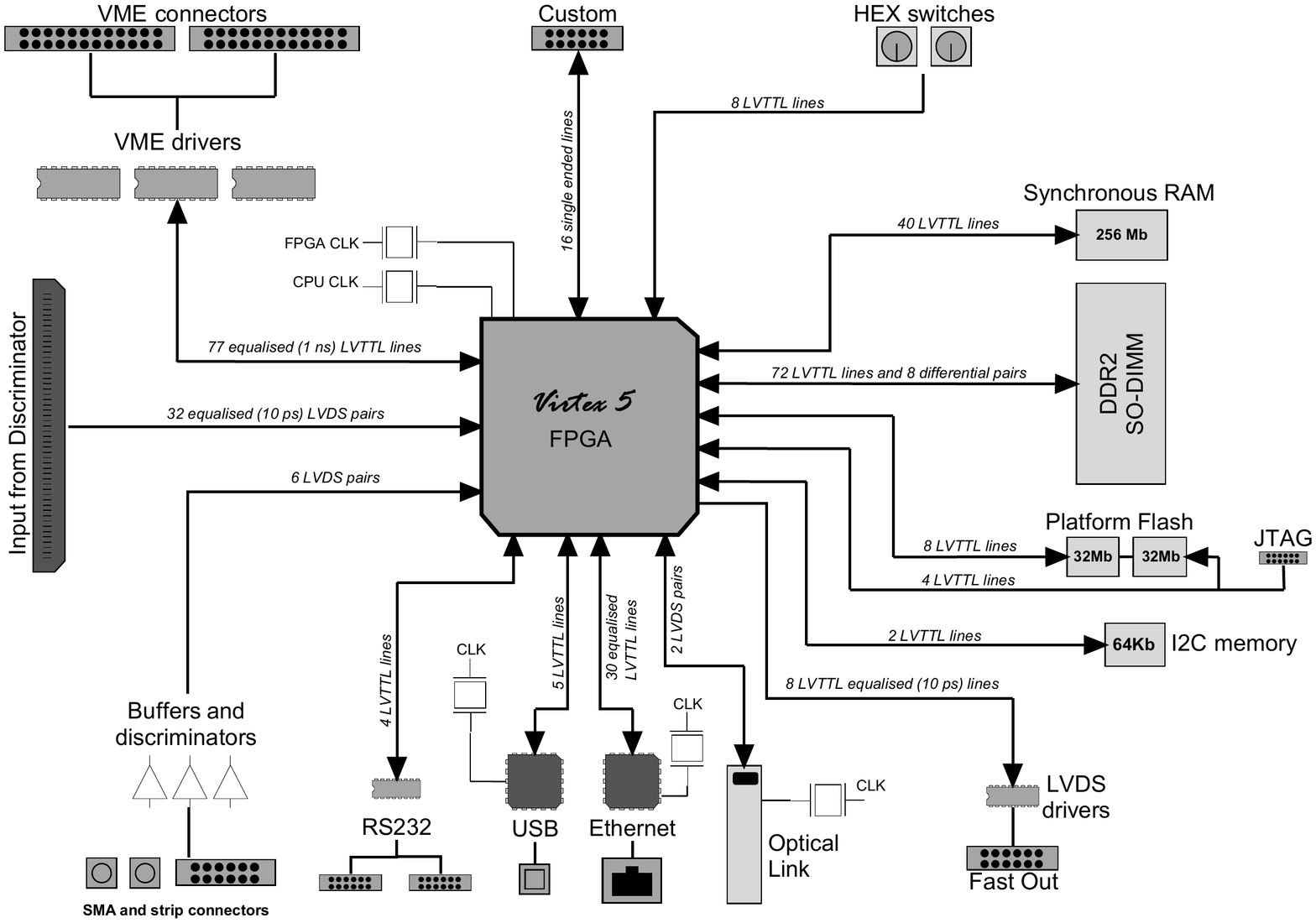}                
		\caption[A chart representing of the HET data acquisition board]{A block diagram of the HET data 
		acquisition board.}             
		\label{fig:v5_acq_sch}                          	
\end{figure}

The HET acquisition system is composed of a set of three electronic boards, and 
one slow control board (see the scheme in Fig. \ref{fig:ELscheme}). The first part 
of the chain, handling PMT analogue signals, is called the front-end electronics 
and is composed of the HET front-end board and the discrimination and shaping 
board. The second part, which performs the measurements and interact with 
signals from KLOE and DA$\Phi$NE, is called HET acquisition system and is 
composed of the HET data acquisition board.

The HET main acquisition is a VME 6U board. The tasks handled by this board are 
to measure the timing of the signals coming from discriminators with respect to DA$\Phi$NE 
fiducial signal, store them only if KLOE trigger are asserted, and transmit data to the KLOE 
acquisition system through the VME bus. To this end, this board is equipped with:

\begin{itemize}
\item Xilinx Virtex 5 FPGA, in which the TDC, the HET DAQ system and the VME
interface are implemented;
\item 6 inputs and outputs to handle DA$\Phi$NE fiducial signal and 
the KLOE trigger signals;
\item 32 LVDS input channels for the TDC;
\item fast VME transceivers;.
\end{itemize}

\begin{figure}[htpb] 
\centering	
		\includegraphics[width= 0.25\textwidth]{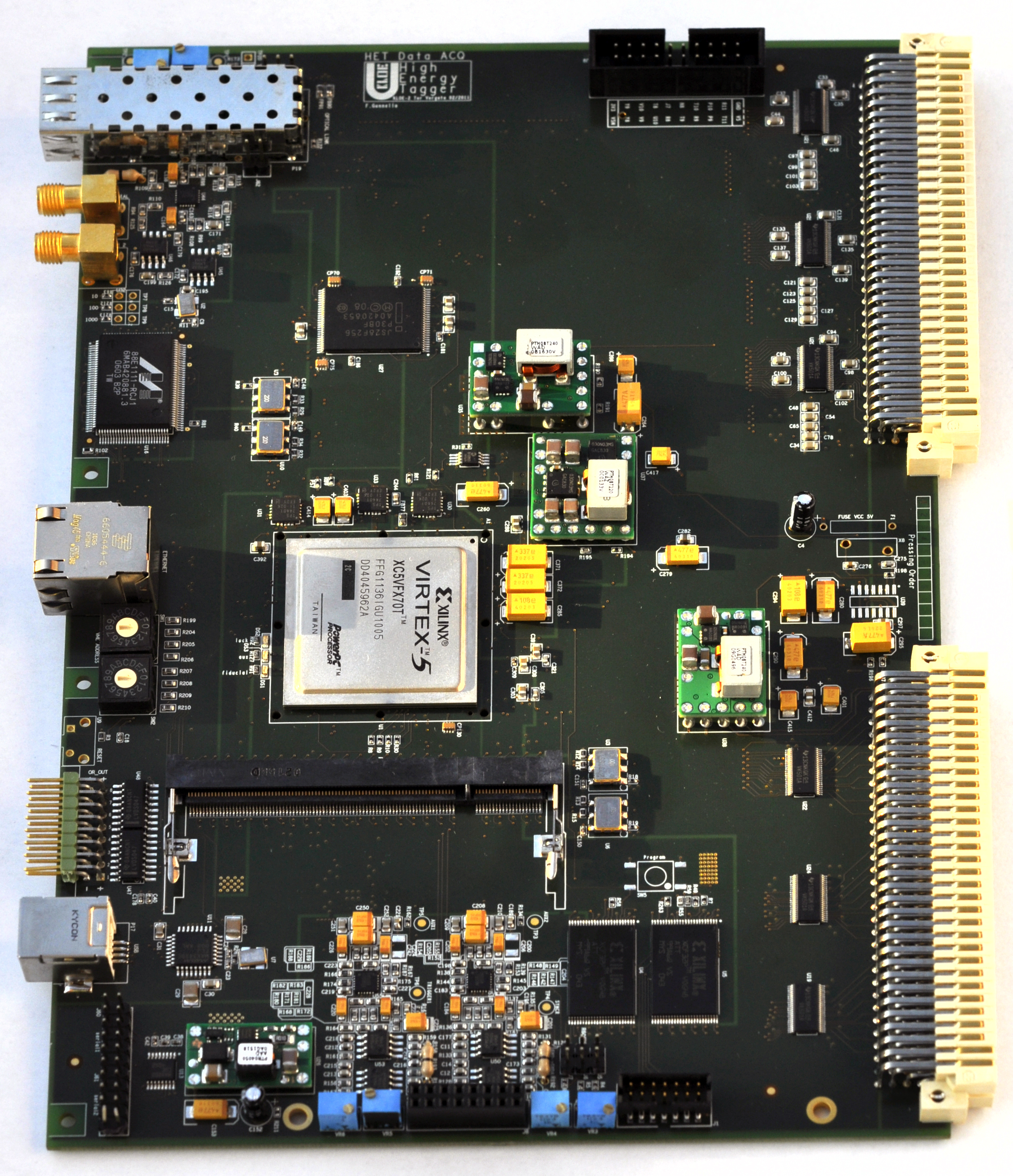}                
		\caption[A picture of the HET data acquisition board]{A picture of the HET data acquisition board.}             
		\label{fig:v5_acq_pic}                          	
\end{figure}

The Virtex 5 XC5VFX70T is the main component of the board, handling all the on-board 
devices (see Fig. \ref{fig:v5_acq_sch} and \ref{fig:v5_acq_pic}). 

%The main features of this 
%FPGA \cite{iafolla}. The highest nominal working frequency of the V5 is \rm{550MHz}, nevertheless the actual 
%working frequency can be sensitively lower, according to the complexity of the implemented logic. 

\section{First Results}
\begin{figure}[htpb] 
\centering	
		\includegraphics[width=.4\textwidth]{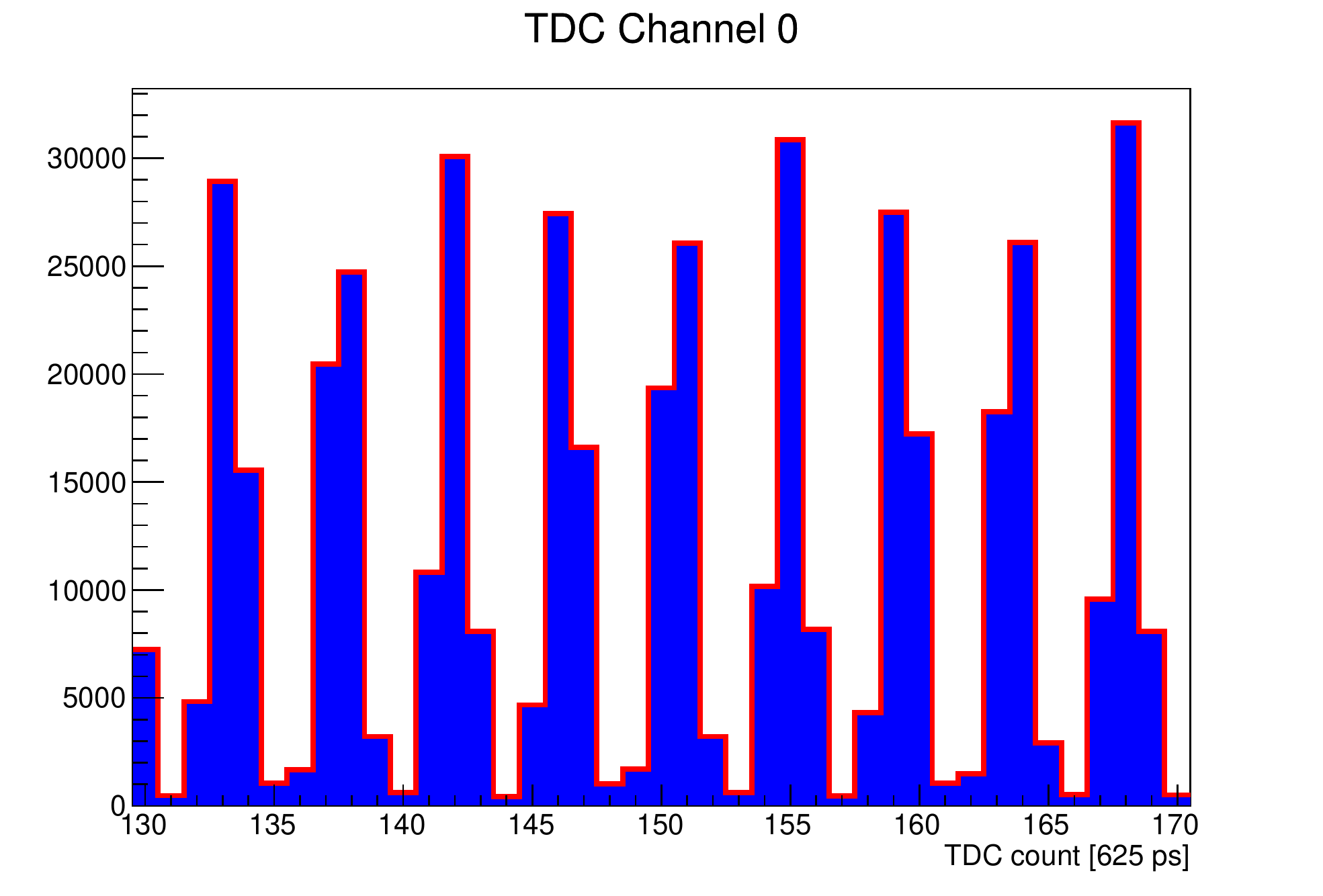} 
		\caption{TDC Spectra for long plastic scintillator.}             
		\label{fig:tdc}                          	
\end{figure}

Both electron and positron arms detectors are now installed and in the commission phase.
Since the bunch crossing occurs in DA$\Phi$NE each $T_{bc} = $~2.7 ns, in order to properly 
disentangle leptons coming from two consecutive bunch crossings, the TDC time resolution 
must be less than $T_{bc}$ as it is shown in Fig. (\ref{fig:tdc}).

%% bibliography

\end{document}